\documentclass[10pt, letterpaper]{article}
\usepackage[colorinlistoftodos]{todonotes}
\usepackage{epsfig}
\usepackage{lscape}
\usepackage{longtable}
\usepackage{amsmath}
\usepackage{amssymb}
\usepackage{amsthm}
\usepackage{setspace}
\usepackage{graphicx}
\usepackage{natbib}
\usepackage{color}
\usepackage{lineno}
\usepackage{subcaption}
\usepackage{framed}
\usepackage{tikz}
\usepackage[labelfont=bf,labelsep=period,justification=justified]{caption}
\usepackage{indent first}
\usepackage{hyperref}
\usepackage[utf8]{inputenc}

\title{{On the number of genealogical ancestors \\ tracing to the source groups of an admixed population}}

\author{Jazlyn A.~Mooney\thanks{Department of Biology, Stanford, CA 94305 USA}  \thanks{Department of Quantitative and Computational Biology, University of Southern California, Los Angeles, CA 90089 USA}, Lily Agranat-Tamir$^*$, Jonathan K.~Pritchard$^{*}$\thanks{Department of Genetics, Stanford University, Stanford, CA 94305 USA}, \\ and Noah A.~Rosenberg$^*$\thanks{Email: noahr@stanford.edu.}}

\date{
{\normalsize  \today}
}

\begin{document}
\maketitle
\normalsize

\fontsize{10.5}{21}\selectfont

\noindent {\bf Abstract.} In genetically admixed populations, admixed individuals possess ancestry from multiple source groups. Studies of human genetic admixture frequently estimate ancestry components corresponding to fractions of individual genomes that trace to specific ancestral populations. However, the same numerical ancestry fraction can represent a wide array of admixture scenarios. Using a mechanistic model of admixture, we characterize admixture genealogically: how many distinct \emph{ancestors from the source populations} does the admixture represent? We consider African Americans, for whom continent-level estimates produce a 75-85\% value for African ancestry on average and 15-25\% for European ancestry. Genetic studies together with key features of African-American demographic history suggest ranges for model parameters. Using the model, we infer that if genealogical lineages of a random African American born during 1960-1965 are traced back until they reach members of source populations, the expected number of genealogical lines terminating with African individuals is 314, and the expected number terminating in Europeans is 51. Across discrete generations, the peak number of African genealogical ancestors occurs for birth cohorts from the early 1700s. The probability exceeds 50\% that at least one European ancestor was born more recently than 1835. Our genealogical perspective can contribute to further understanding the admixture processes that underlie admixed populations. For African Americans, the results provide insight both on how many of the ancestors of a typical African American might have been forcibly displaced in the Transatlantic Slave Trade and on how many separate European admixture events might exist in a typical African-American genealogy.

\fontsize{10.5}{21}\selectfont
\clearpage

\section*{Introduction}
\label{secIntroduction}

\noindent Genetically admixed populations arise when two or more source groups combine to form a new population. After a period of multiple generations of mating among members of the incipient admixed population and new contributors from the source groups, typical individuals in the admixed population possess ancestry from multiple sources~\citep{Chakraborty86, KorunesAndGoldberg21,GopalanEtAl22}.

The genetic history of an admixed population can be represented by a temporal sequence of admixture contributions, starting with the founding of the new admixed group~\citep{Long91, VerduAndRosenberg11, Gravel12}. Among present-day members of the admixed population, genetic patterns such as the distribution of admixture levels estimated from individual genomes can then be used together with a model of the admixture process to uncover features such as the timing and magnitude of the genetic contributions that characterize the admixture process~\citep{VerduEtAl14, BaharianEtAl16, ZaitlenEtAl17}.

In studies that seek to infer population parameters from genetic patterns among individuals in the admixed population, each admixed individual is treated as a random outcome of the admixture process. The accumulation of data on many admixed individuals then provides information about the population history. In this perspective, for a given model of the admixture history, an individual possesses a random genealogy conditional on the parameters of the admixture process. What information can be obtained about a random individual genealogy under the assumptions of an admixture model? In particular, for individual members of an admixed population, how many distinct contributors from the source populations does their admixture represent?

In human admixed populations, questions focused on random individual genealogies can provide information both about the population-level history of admixture and about the relationship of individuals to that history. Consider the case of the African-American admixed population in the United States. Living African Americans descend primarily from an admixture of African and European source populations, much of the admixture having occurred during the period of enslavement of most African Americans, 1619-1865. Owing to widespread patterns such as forcible fracturing of enslaved families by enslavers, a practice of using only first names and not surnames for enslaved persons, lack of documentation of many of the enslaved even by first name in the written record, and a reticence of many formerly enslaved individuals to record genealogical information in the period after slavery, for many African Americans, limited data are available about their individual ancestors prior to the middle or late 1800s~\citep{Gates09, Swarns12, Nelson16}. Thus, an admixture model has potential to recover features of African-American genealogies that are otherwise difficult to obtain.

For an African American chosen at random, how many genealogical lines traced back from the present to a member of a source population reach an African individual? How many reach a European or European American? The former quantity approximates the number of ancestors who traveled from Africa to the Western Hemisphere as forced enslaved migrants in the Transatlantic Slave Trade. The latter gives the number of occasions at which European admixture events occurred in a random African-American genealogy. Answers to such questions are informative not only for understanding the genealogies of individuals, but also for contributing details of the admixture process that has given rise to the present-day population.

\section*{Model}

\subsection*{Assumptions}

\begin{figure}[tb]
\centerline{\includegraphics[width=0.8\textwidth]{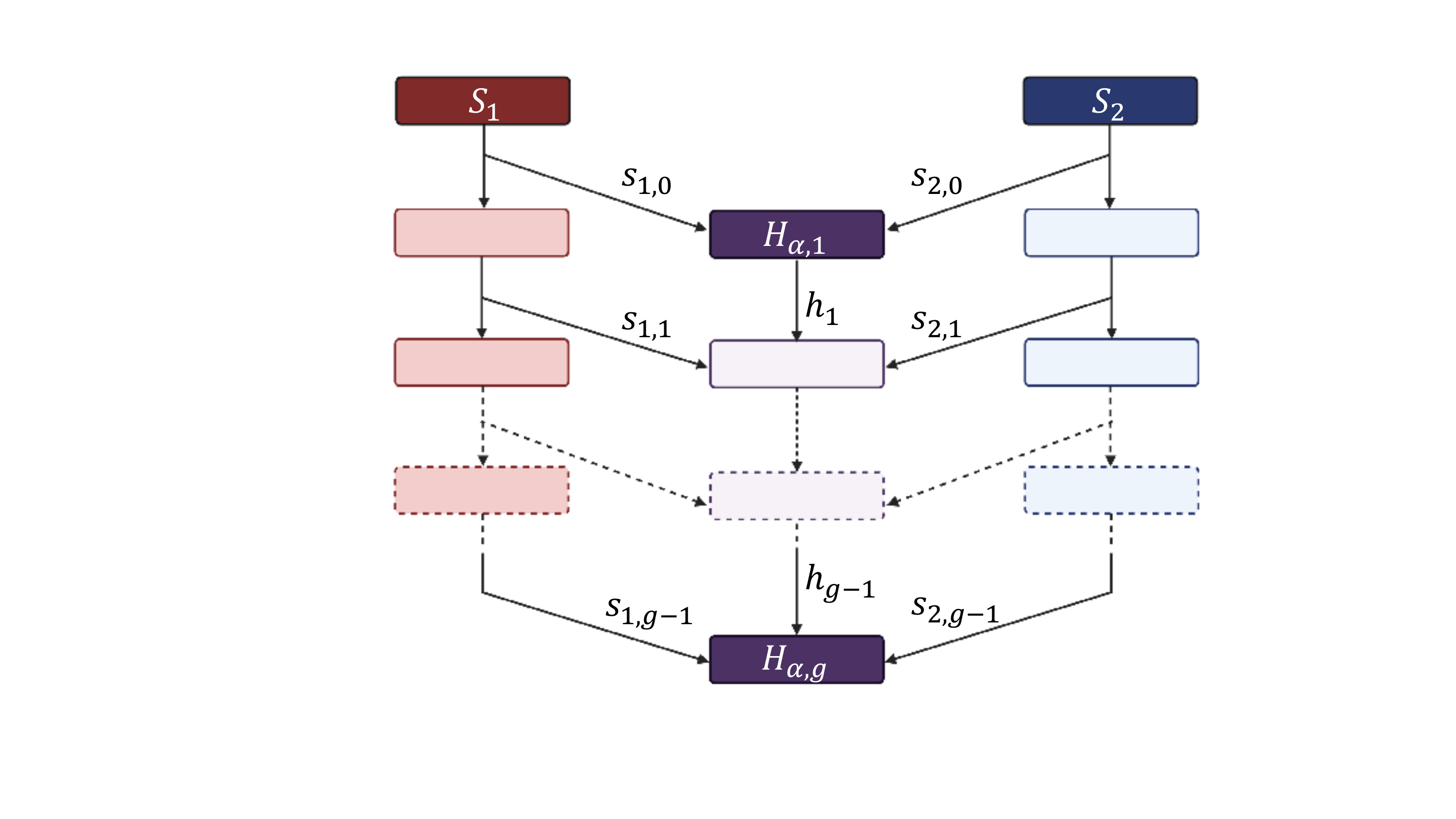}}
\vspace{-0.25cm}
\caption{Schematic of the admixture model. Source populations $S_1$ and $S_2$ contribute to an admixed population $H$. The members of $H$ in generation $g$ draw parents from the populations of generation $g-1$ from $S_1$ with probability $s_{1,g-1}$, from $H$ with probability $h_{g-1}$, and from $S_2$ with probability $s_{2,g-1}$. Two parents are drawn independently. Random variable $H_{\alpha,g}$ denotes the random autosomal ancestry fraction from population $\alpha$ (1 for $S_1$, 2 for $S_2$) in an individual in population $H$ in generation $g$.}\label{fig1}
\end{figure}

\noindent We follow a mechanistic model in which admixture levels are explored in an admixed population over time~\citep{VerduAndRosenberg11, GoldbergEtAl14, GoldbergAndRosenberg15, GoldbergEtAl20}. Three populations are considered: source populations $S_1$ and $S_2$, and admixed population $H$. In each of a series of generations---indexed discretely with the index increasing forward in time---an individual in the admixed population $H$ in generation $g$ has a pair of parents probabilistically drawn from among individuals extant in generation $g-1$ in source populations $S_1$ and $S_2$ and admixed population $H$ (Figure \ref{fig1}).

Suppose that for an individual in generation $g$, the admixture contributions are $s_{1,g-1}$, $s_{2,g-1}$, and $h_{g-1}$, for populations $S_1$, $S_2$, and $H$, respectively. In other words, for an individual chosen at random in admixed population $H$, a parent chosen at random has probability $s_{1,g-1}$ of having originated from population $S_1$, $s_{2,g-1}$ for population $S_2$, and $h_{g-1}$ for population $H$. We then have
\begin{eqnarray}
\label{eqSumOfF}
s_{1,g-1} + h_{g-1} + s_{2,g-1} & = & 1.
\end{eqnarray}
The sampling probabilities $s_{1,g-1}$, $s_{2,g-1}$, and $h_{g-1}$ can be interpreted as fractional contributions from source populations $S_1$, $S_2$, and $H$ to autosomal genomes in population $H$ in generation $g$. Generation $g=1$ represents the founding of the admixed population from members of the source population from generation $g=0$. The admixed population does not exist in generation $g=0$, so that $h_0=0$, and $s_{1,0}+s_{2,0}=1$.

\begin{figure}[tb]
\centerline{\includegraphics[width=\textwidth]{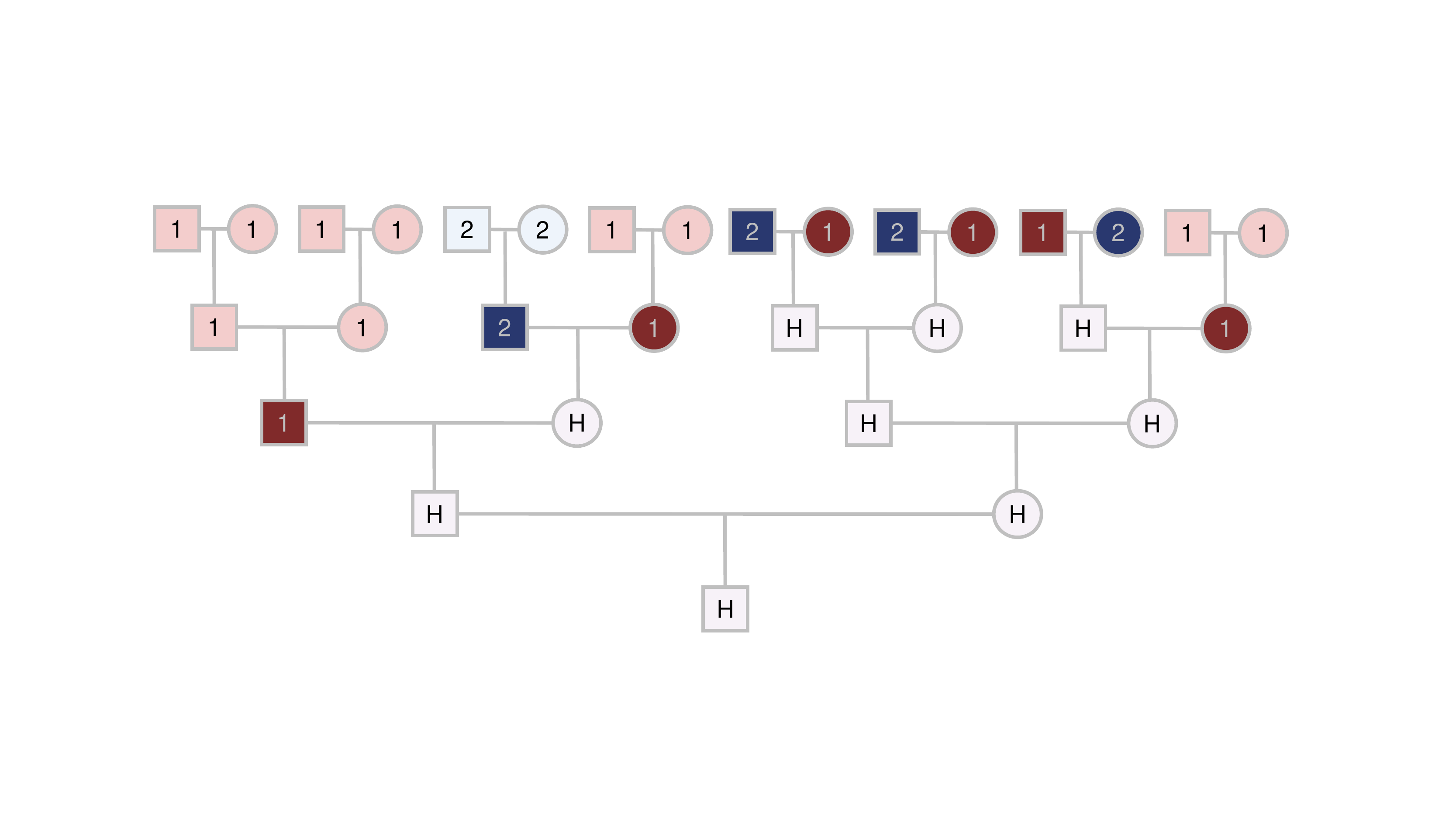}}
\vspace{-1.5cm}
\caption{Counting genealogical ancestors. The pedigree of the individual at the bottom of the diagram is traced back in time until ancestral populations are reached. Each individual in the pedigree is labeled by the population to which it belongs: source population 1 (red), source population 2 (blue), or admixed population H (purple). For the index individual, this pedigree shows six ancestors from source 1 and four from source 2. The count of genealogical ancestors from the source populations tabulates, along each ancestral line, the \emph{first} individual reached who belongs to a source population: the six individuals from source 1 shown in dark red and the four individuals from source 2 shown in dark blue. The admixture fractions for the individual are $\frac{11}{16}$ from source 1 and $\frac{5}{16}$ from source 2.}\label{fig2}
\end{figure}

Previous studies with these modeling assumptions have tracked properties of random variables that describe \emph{admixture proportions} in the source populations $S_1$ and $S_2$ at generation $g$.  In particular, \cite{VerduAndRosenberg11} studied recursions for the probability distribution and moments of a random variable $H_{1,g}$, representing the autosomal fraction of admixture from source population 1 for an individual in the admixed population at generation $g$. We instead study the random variable $Z_{1,g}$, the \emph{number of genealogical ancestors} from source population 1 for an individual in the admixed population at generation $g$, and $Z_{2,g}$, the number of genealogical ancestors from source population 2. In the sense in which we consider genealogical ancestors, once a source population is reached along a genealogical line in a specific ancestor, that ancestor is tabulated as a genealogical ancestor from the associated source population, and the line is not traced any farther back (Figure \ref{fig2}).

\subsection*{Recursion for the number of genealogical ancestors}

\noindent We review expressions that we will need for the mean and variance of autosomal admixture under the model~\citep{VerduAndRosenberg11}. The mean ancestry fraction from population 1 in generation $g$ is~\citep[][eqs.~10 and 11]{VerduAndRosenberg11}:
\begin{equation}
\label{eqMeanReview}
\mathbb{E}[H_{1,g}]=
\begin{cases}
s_{1,0}, & g=1, \\
s_{1,g-1} + h_{g-1} \mathbb{E}[H_{1,g-1}], & g \geq 2.
\end{cases}
\end{equation}
The variance of the ancestry fraction from population 1 in generation $g$ is~\citep[][eqs.~22 and 23]{VerduAndRosenberg11}
\begin{align}
\label{eqVarReview}
\mathbb{V}[H_{1,g}]=
\begin{cases}
\frac{s_{1,0}(1-s_{1,0})}{2}, & g=1, \\
\begin{aligned}[b]
\tfrac{s_{1,g-1}(1-s_{1,g-1})}{2} - s_{1,g-1}h_{g-1} \mathbb{E}[H_{1,g-1}] + \tfrac{h_{g-1}(1-h_{g-1})}{2} \mathbb{E}\left[H_{1,g-1} \right]^2 \\
 +\tfrac{h_{g-1}}{2}\mathbb{V}[H_{1,g-1}],
\end{aligned} & g \geq 2. \\
\end{cases}
\end{align}
Note that the mean ancestry fraction from population 2 is one minus the mean ancestry fraction from population 1, and the variances of the two ancestry fractions are equal.

A recursion describing the admixture fraction $H_{1,g}$~\citep{VerduAndRosenberg11} can be modified to obtain a recursion for $Z_{1,g}$. Whereas the random autosomal admixture fraction $H_{1,g}$ of an individual is the mean of the corresponding admixture fractions of the parents of the individual, the random number of ancestors $Z_{1,g}$ is the \emph{sum} of the numbers of ancestors of the parents (from population 1).

Let $L$ be a random variable that gives the source populations of the parents of a random individual from the admixed population. Listing the mother first, $L$ takes a value in the set $\mathcal{L}=\{S_1S_1, S_1H, S_1S_2, HS_1, HH, HS_2, S_2S_1, S_2H, S_2S_2\}$. Based on eqs.~1 and 2 of \cite{VerduAndRosenberg11}, for generation $g=1$, we have
\begin{equation}
\label{eqZRecursionBaseCase}
Z_{1,1} =
\begin{cases}
2 & \text{if } L=S_1S_1, \text{with } \mathbb{P}[L=S_1S_1]=s_{1,0}s_{1,0} \\
1 & \text{if } L=S_1S_2, \text{with } \mathbb{P}[L=S_1S_2]=s_{1,0}s_{2,0} \\
1 & \text{if } L=S_2S_1, \text{with } \mathbb{P}[L=S_2S_1]=s_{2,0}s_{1,0} \\
0 & \text{if } L=S_2S_2, \text{with } \mathbb{P}[L=S_2S_2]=s_{2,0}s_{2,0}.
\end{cases}
\end{equation}
For subsequent generations, $g \geq 2$,
\begin{equation}
\label{eqZRecursion}
Z_{1,g} =
\begin{cases}
2             & \text{if } L=S_1S_1, \text{with } \mathbb{P}[L=S_1S_1]=s_{1,g-1}s_{1,g-1} \\
1+Z_{1,g-1}   & \text{if } L=S_1H,   \text{with } \mathbb{P}[L=S_1H]=s_{1,g-1}h_{g-1} \\
1             & \text{if } L=S_1S_2, \text{with } \mathbb{P}[L=S_1S_2]=s_{1,g-1}s_{2,g-1} \\
Z_{1,g-1}+1   & \text{if } L=HS_1,   \text{with } \mathbb{P}[L=HS_1]=h_{g-1}s_{1,g-1} \\
Z_{1,g-1}+Z^\prime_{1,g-1} & \text{if } L=HH, \text{with } \mathbb{P}[L=HS_1]=h_{g-1}h_{g-1} \\
Z_{1,g-1}     & \text{if } L=HS_2,   \text{with } \mathbb{P}[L=HS_2]=h_{g-1}s_{2,g-1} \\
1             & \text{if } L=S_2S_1, \text{with } \mathbb{P}[L=S_2S_1]=s_{2,g-1}s_{1,g-1} \\
Z_{1,g-1}     & \text{if } L=S_2H,   \text{with } \mathbb{P}[L=S_2H]=s_{2,g-1}h_{g-1} \\
0             & \text{if } L=S_2S_2, \text{with } \mathbb{P}[L=S_2S_2]=s_{2,g-1}s_{2,g-1}.
\end{cases}
\end{equation}

For $L=HH$, $Z_{1,g-1}$ and $Z_{1,g-1}^\prime$ are independent and identically distributed copies of the same random variable. Eqs.~\ref{eqZRecursionBaseCase} and \ref{eqZRecursion} enable us to compute the probability distribution of $Z_{1,g}$, the number of population-1 ancestors of an individual in the admixed population in generation $g$. $Z_{1,g}$ and $Z_{2,g}$ range in $Q_g = \{0, 1, \ldots, 2^g\}$. For $q$ in $Q_g$, we compute the probability $\mathbb{P}[Z_{1,g}=q]$ that a random individual from population $H$ at generation $g$ has $q$ genealogical ancestors from population 1.

Analogously to eqs.~3-5 of~\cite{VerduAndRosenberg11}, we have for $g \geq 1$
\begin{equation}
\label{eqZDistributionBaseCase}
\mathbb{P}[Z_{1,1}=q] =
\begin{cases}
s_{1,0}^2,        & q=2, \\
2s_{1,0}s_{2,0},  & q=1, \\
s_{2,0}^2,        & q=0.
\end{cases}
\end{equation}
For $g \geq 2$ and $q$ in $Q_g$,
\begin{eqnarray}
\label{eqZDistribution}
\mathbb{P}[Z_{1,g}=q] & = & h_{g-1}^2 \displaystyle\sum_{r=0}^{2^{g-1}} \big( \mathbb{P}[Z_{1,g-1} = r] \, \mathbb{P}[Z_{1,g-1}=q-r] \big) \nonumber \\
& & + (2s_{1,g-1}h_{g-1}) \, \mathbb{P}[Z_{1,g-1}=q-1] + (2s_{2,g-1}h_{g-1})  \, \mathbb{P}[Z_{1,g-1}=q] + I_g(q).
\end{eqnarray}
Function $I_g$ is equal to
\begin{equation}
\label{eqIgq}
I_g(q) =
\begin{cases}
s_{1,g-1}^2,             & q=2, \\
2s_{1,g-1}s_{2,g-1}      & q=1, \\
s_{2,g-1}^2,             & q=0, \\
0,                       & 3 \leq q \leq 2^q.
\end{cases}
\end{equation}
Eq.~\ref{eqZDistribution} sums over all possible parental pairings that lead to $q$ ancestors from population 1 at generation $g$. Only three values of $q$ are possible if neither parent is from the admixed population---$q=0$, $q=1$, and $q=2$---producing the terms in eq.~\ref{eqIgq}.

\subsection*{Recursive mean and variance of the number of genealogical ancestors}
\noindent Using the recursion for the probability distribution of the number of ancestors in eqs.~\ref{eqZRecursionBaseCase} and \ref{eqZRecursion}, we follow \cite{VerduAndRosenberg11} to obtain the moments of $Z_{1,g}$. By the law of conditional expectation,
\begin{equation}
\mathbb{E}[Z_{1,g}]=
\mathbb{E}_L\left[\mathbb{E}[Z_{1,g}|L]\right] = \sum_{\ell \in \mathcal{L}}
\mathbb{P}[L=\ell] \, \mathbb{E}[Z_{1,g}|L=\ell].
\end{equation}
For each $\ell \in \mathcal{L}$,  $\mathbb{E}[Z_{1,g}|L=\ell] = 2\mathbb{E}[H_{1,g}|L=\ell]$, so that the recursive computation of $\mathbb{E}[Z_{1,g}]$ follows that of $\mathbb{E}[H_{1,g}]$ in eqs.~6-11 of \cite{VerduAndRosenberg11}, multiplying by a factor of 2. We obtain $\mathbb{E}[Z_{1,g}] = 2\mathbb{E}[H_{1,g}]$, or
\begin{equation}
\label{eqMean}
\mathbb{E}[Z_{1,g}]=
\begin{cases}
2s_{1,0}, & g=1, \\
2s_{1,g-1}+2h_{g-1}\mathbb{E}[Z_{1,g-1}], & g \geq 2.
\end{cases}
\end{equation}

For the $k$th moment of $Z_{1,g}$, for each $\ell$, $\mathbb{E}[Z_{1,g}^k|L=\ell] = 2^k\mathbb{E}[H_{1,g}^k|L=\ell]$. In particular, as $\mathbb{E}[Z_{1,g}^2|L=\ell] = 4\mathbb{E}[H_{1,g}^2|L=\ell]$, we obtain $\mathbb{E}[Z_{1,g}^2] = 4\mathbb{E}[H_{1,g}^2]$. Because $\mathbb{E}[Z_{1,g}]^2 = 4\mathbb{E}[H_{1,g}]^2$ and $\mathbb{E}[Z_{1,g}^2] = 4\mathbb{E}[H_{1,g}^2]$, we have $\mathbb{V}[Z_{1,g}]=4\mathbb{V}[H_{1,g}]$. We apply eqs.~22 and 23 of \cite{VerduAndRosenberg11} for $\mathbb{V}[H_{1,g}]$, obtaining
\begin{equation}
\label{eqVar}
\mathbb{V}[Z_{1,g}]=
\begin{cases}
2s_{1,0}(1-s_{1,0}),         & g=1, \\
\begin{aligned}[b]
2s_{1,g-1}(1-s_{1,g-1}) - 4s_{1,g-1}h_{g-1} & \mathbb{E}[Z_{1,g-1}] \\
+ 2h_{g-1}(1-h_{g-1}) & \mathbb{E}\left[Z_{1,g-1} \right]^2 \\
+ 2h_{g-1}            & \mathbb{V}[Z_{1,g-1}],
\end{aligned}         & g \geq 2.
\end{cases}
\end{equation}

To obtain $\mathbb{P}[Z_{2,g}=q]$, $\mathbb{E}[Z_{2,g}]$, and $\mathbb{V}[Z_{2,g}]$, we substitute analogous quantities $s_{2,0}$ and $s_{2,g-1}$ in place of the quantities $s_{1,0}$ and $s_{1,g-1}$ used to produce $\mathbb{P}[Z_{1,g}=q]$, $\mathbb{E}[Z_{1,g}]$, and $\mathbb{V}[Z_{1,g}]$ in eqs.~\ref{eqZRecursionBaseCase}-\ref{eqVar}.

\subsection*{Nonrecursive mean number of genealogical ancestors}

\noindent A nonrecursive solution for the mean number of genealogical ancestors from population 1, $\mathbb{E}[Z_{1,g}]$, can be obtained from  eq.~\ref{eqMean}. Iterating eq.~\ref{eqMean} from generation $g$ back to generation 0, we have
\begin{equation}
\label{eqNonrecursive}
\mathbb{E}[Z_{1,g}] = \sum_{i=0}^{g-1} \bigg( 2s_{1,i} \prod_{j=i+1}^{g-1} 2h_j \bigg), \, g \geq 1.
\end{equation}
The sum in eq.~\ref{eqNonrecursive} decomposes the expression for $\mathbb{E}[Z_{1,g}]$ into terms that represent ancestors from specific generations. The expected number of genealogical ancestors in generation $g$ is a sum of values contributed by generations $0,1,\ldots,g-1$. In particular, the summand $2s_{1,i} \prod_{j=i+1}^{g-1} 2h_j$ represents the expected number of genealogical ancestors contributed by generation $i$, $0 \leq i \leq g-1$, to a randomly chosen individual living in the admixed population in generation $g$. A similar nonrecursive expression can be obtained for  $\mathbb{E}[Z_{2,g}]$, substituting $s_{2,i}$ in place of $s_{1,i}$.

\subsection*{Probability of at least one genealogical ancestor in a specified generation}

\noindent The model also enables a calculation of the probability that an individual from the admixed population has \emph{at least one} genealogical line terminating in a specified source population in a specified generation. For $n=0,1,\ldots,g$, let $X_n$ denote, for an individual in the admixed population $H$ in generation $g$, the individual's number of ancestors in generation $n$ who are also in $H$. Then $X_g=1$, and for each $n=0,1,\ldots,g-1$, $X_n$ is a random variable ranging in $[0,2^{g-n}]$. As each of the 2 parents of a random individual from generation $n+1$ is a Bernoulli trial with probability $h_n$ of being from the admixed population, $X_n$ is recursively distributed as $X_n \sim \text{Bin} (2X_{n+1}, h_n)$.

For each $n=0,1,\ldots,g-1$, denote by $U_n$ the random number of ancestors that an individual in population $H$ in generation $g$ has in population $S_1$ in generation $n$. This quantity is the number of ancestral lines of a random member of $H$ in generation $g$ that reach $S_1$ precisely in generation $n$. For each $n=0,1,\ldots,g-1$, if $X_{n+1}=0$, then $U_n=0$; otherwise $U_n \sim \text{Bin} (2X_{n+1}, s_{1,n})$.

For $n=0,1,\ldots,g-1$, we compute $1-\mathbb{P}[U_n=0]$, the probability that a random admixed individual has at least one ancestral line that reaches population $S_1$ in generation $n$. By the law of total probability,
\begin{eqnarray}
\mathbb{P}[U_n=0] & = & \sum_{m=0}^{2^{g-(n+1)}} \mathbb{P}[U_n=0 | X_{n+1} = m] \, \mathbb{P}[X_{n+1}=m] \nonumber \\
                  & = & \mathbb{P}[X_{n+1}=0] + \sum_{m=1}^{2^{g-(n+1)}} (1-s_{1,n})^{2m}  \, \mathbb{P}[X_{n+1}=m]. \label{eq:U}
\end{eqnarray}
After recursively computing $\mathbb{P}[X_{n}=m]$ for each $n=g-1,g-2,\ldots,0$ for all $m=0,1,\ldots,2^{g-n}$, eq.~\ref{eq:U} can be evaluated as a function of the parameters $s_{1,n}$ and $h_n$ for $n=0,1,\ldots,g-1$. We then obtain the desired probability $1-\mathbb{P}[U_n=0]$ for each $n$. A similar calculation can evaluate the probability that a random member of the admixed population has at last one ancestral line terminating in population $S_2$ in generation $n$; we simply substitute $s_{2,n}$ in place of $s_{1,n}$.

\section*{Application to African-American genealogies}

\subsection*{Overview of the model for African-American admixture history}

\noindent We use the admixture model to count genealogical ancestors for individuals chosen at random in the African-American population. Our approach involves fitting the model to data on African-American genetic ancestry. We thus estimate admixture parameters under the model, obtaining the expected numbers of African and European genealogical ancestors as byproducts of the estimation.

We constrain the model by using known features of African-American demographic history~\citep{Berlin10, EltisAndRichardson10, FranklinAndHigginbotham21}. Starting from the founding of the African-American population, the admixture history of the population can be divided into three demographic epochs prior to 1965: 1619-1808, 1808-1865, and 1865-1965. In the first period, the population was formed from African and European sources, with both sources contributing to the emerging admixed population throughout the period. In the second period, with the end of legal importation of enslaved African captives into the United States, contributions from the African source were much reduced, with contributions from Europeans and European Americans continuing. In the third period, the end of legal enslavement, and hence of its accompanying forms of coercive mating between enslavers and enslaved persons, may have reduced contributions from the European and European-American source, with contributions from the African source remaining low. A three-epoch admixture model for births prior to 1965 accords with genetic evidence supporting such a division, with dates similar to those suggested by historical periods~\citep{BaharianEtAl16}.

We focus our attention on the birth cohort 1960-1965 as an endpoint for the model. This cohort is sensible first because much of the genetic data from which model parameters can be estimated traces largely to studies of adult diseases, representing individuals born approximately in this time period. Second, the period after 1965 would introduce a demographically distinct fourth epoch---with additional parameters to estimate---as African-American births after 1965 reflect increased contributions of the African source after an increase in African immigration, and increased contributions of the European source after relaxations of laws and norms limiting acceptance of unions between Africans or African Americans and Europeans or European Americans.

\begin{figure}[tb]
\begin{center}
\centerline{\includegraphics[width=0.5\textwidth]{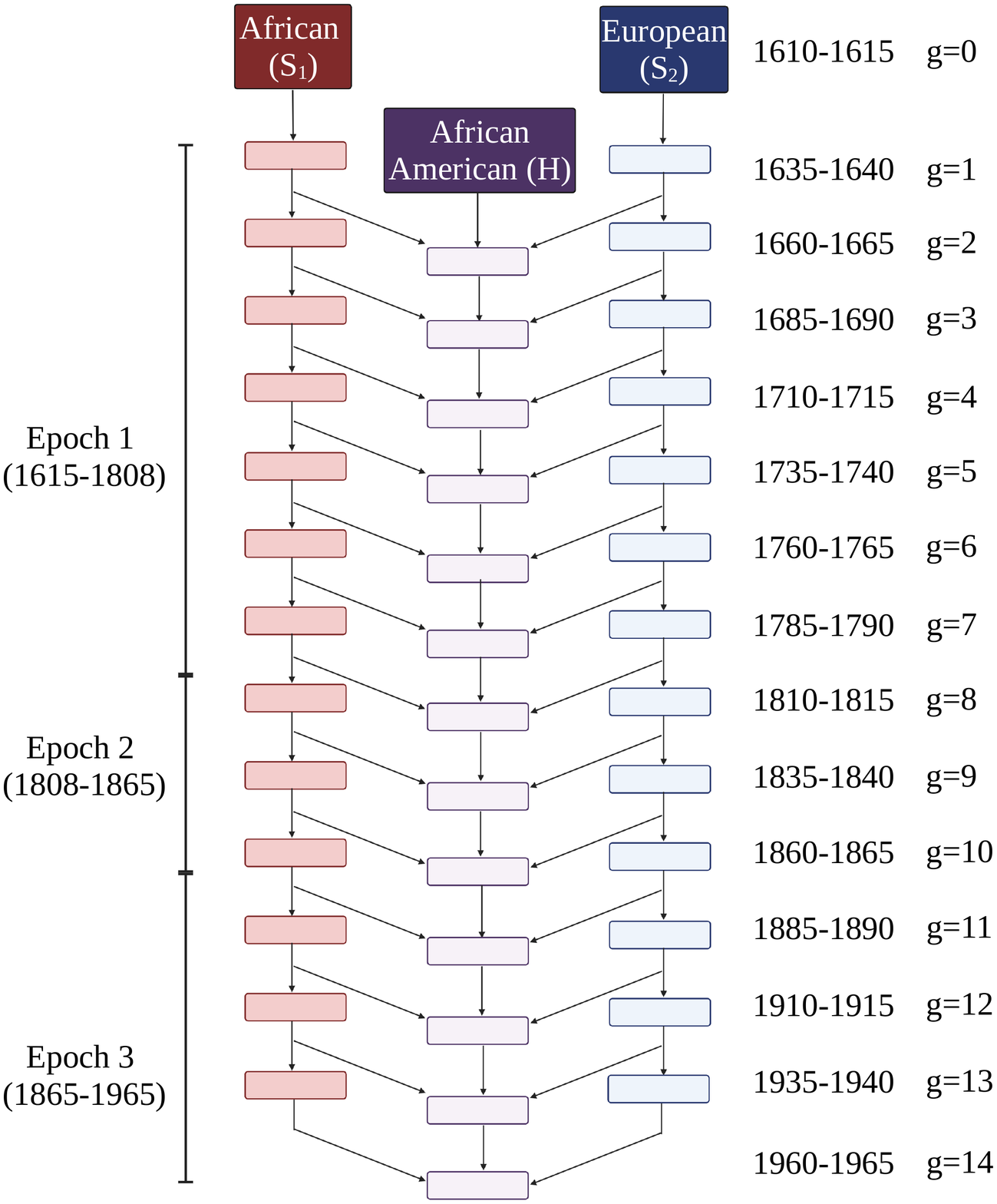}}
\end{center}
\vspace{-0.4cm}
\caption{The admixture model for African Americans. The model is a special case of Figure \ref{fig1}. $S_1$ denotes Africans, $S_2$ denotes Europeans and European Americans, and $H$ denotes African Americans. We consider the births in a 5-year interval to be a discrete generation $g$, with $g=0$ corresponding to 1610-1615 and $g=14$ to 1960-1965, and we assume a 25-year generation time. The model has three epochs, with epochs 1, 2, and 3 corresponding to generations 1-7, 8-10, and 11-14, respectively.}
\label{fig3}
\end{figure}

With a 25-year generation time, the third epoch contains four generational birth cohorts (1885-1890, 1910-1915, 1935-1940, 1960-1965), the second epoch has three, and the first has seven. Thus, the model has $g=14$ generations, with generation 14 born during 1960-1965 (Figure \ref{fig3}).

In our application of the model, we note a subtle aspect of the meaning we use for ``African genealogical ancestors.'' A person born in Africa who arrived in North America is regarded as ``African''; in counting African ancestors, we count African migrants in the ancestry of an African American. All births in the model take place in the admixed population in North America; a person born in this admixed population is regarded as an ``African American.'' It is possible for an African American in the model to have all genealogical ancestors from Africa (or, in principle, from Europe, though this scenario is unlikely in the relevant portion of the parameter space). Irrespective of the person's genetic ancestry, however, such a person is regarded as an African American.

The approach treats ``European and European-American'' genealogical ancestors as a single population category, not distinguishing between individuals born in Europe and those born in North America. For simplicity, we abbreviate this population as ``European.''

\subsection*{Constraining the three-epoch model by demographic data}

\noindent Without loss of generality, we treat the African source population as population 1 and the European source population as population 2. We set $s_{1,0}=1$ and $s_{2,0}=0$, founding the African-American population with Africans in the first generation $g=1$. In the three-epoch model, after the founding with births in generation 1 to parents from generation 0, matings occur intragenerationally between members of generations 1-6 in epoch 1, 7-9 in epoch 2, and 10-13 in epoch 3.

We make use of demographic data to initialize the model for the duration of the first epoch~\citep{Hacker20}. At the start of epoch 1, an individual born in the admixed African-American population in generation 1 has parents only from the African and European populations, and not from the African-American population---as the African-American population did not yet exist in the parental generation 0 (we further assume that all parents of individuals in generation 1 are African). By the end of this epoch, an individual born in the admixed African-American population has a high probability of having one or both parents from the African-American population, as the size of the African-American population had grown to exceed the number of arriving Africans.

Let $c_{g-1} = s_{1,g-1} / (s_{1,g-1} + h_{g-1}) = s_{1,{g-1}} / (1-s_{2,{g-1}})$, denoting, for individuals born in the African-American population in generation $g$, the fraction of their non-European parents who are African arrivals to North America rather than African-American residents. We assume that these parents are drawn in proportion to the population sizes of potential African and African-American parents available at the time of the birth of generation $g$. Hence, we write $c_{g-1} = \mathcal{S}_{1,g} / (\mathcal{S}_{1,g} + \mathcal{H}_{g-1})$, where $\mathcal{S}_{1,g}$ is the estimated number of African arrivals in generation $g$, entrants assumed to be of child-bearing age and hence potential parents of individuals born in generation $g$, and $\mathcal{H}$ is the number of births in the African-American population in generation $g-1$, members of the previous generation who are also potential parents of individuals born in generation $g$.

To choose values for $c_{g-1}$, we use estimated numbers of migrants and births from demographic analysis of the enslaved population~\citep[][columns 7 and 8 of Table 1]{Hacker20}. Each of our generations is a 5-year interval; we use data reported for the corresponding 10-year interval of which that 5-year interval is a sub-interval. Thus, for example, $c_2$, representing the fraction of non-European parents of African Americans born in generation 3 (1685-1690) who are African, is the ratio of the estimated number of African migrants in 1680-1690 to the sum of this quantity and the estimated number of African-American births 1660-1670 (representing generation 2, 1660-1665). Note that the demographic study~\citep{Hacker20} focuses on enslaved Africans and African Americans; we assume that its demographic parameters apply to the entire population of Africans and African Americans.

With this approach, in epoch 1, for each generation 1 to 7, we seek to estimate the model parameters $(s_{1,g-1},h_{g-1},s_{2,g-1})$ subject to the constraints that for each $g$ from 1 to 7, $s_{1,g-1}=c_{g-1} (1-s_{2,{g-1}})$ and $h_{g-1}=(1-c_{g-1})(1-s_{2,g-1})$, with each $c_{g-1}$ fixed according to the entries of Table \ref{table:1cg} and with $s_{2,g-1}$ equal to the same value for each $g$ from 1 to 7 (to be precise, note that for $g=1$, no estimation is needed, as $s_{1,0}$ is fixed at 1). In the more recent epochs 2 and 3, we estimate all model parameters $(s_{1,g-1}, h_{g-1}, s_{2,g-1})$ associated with births in generation $g$, without such constraints. Across the generations within epochs 2 and 3, we assume parameter values are constant, and we index parameters by the first of the contributing generations: 7 and 10. Thus, model parameters for these epochs are $(s_{1,7},h_{7},s_{2,7})$ and $(s_{1,10},h_{10},s_{2,10})$, with only two of each parameter trio being free to vary, and the third equaling one minus the sum of the other two (eq.~\ref{eqSumOfF}). Because model parameters are constant within epochs 2 and 3, we treat model parameters as equal across generations in the recursions that give rise to generations 8 to 10 and in those that give rise to generations 11 to 14.

\subsection*{Fitting the model}

\noindent To fit the model, we search the parameter space, for each choice of model parameters computing the mean and variance of autosomal admixture in generation $g=14$. We compute $\mathbb{E}[H_{1,14}]$ and $\mathbb{V}[H_{1,14}]$ by recursively applying eqs.~\ref{eqMeanReview} and \ref{eqVarReview}; we proceed similarly for $\mathbb{E}[H_{2,14}]$ and $\mathbb{V}[H_{2,14}]$.

Estimates of African and European ancestry in studies of African-American admixture in different locations and in different conditions of health and disease have been generally concordant, with values of $\sim$80\% for the mean African ancestry and $\sim$10\% for the standard deviation. For example, in 14 data sets on African-American admixture tabulated by~\cite{ChengEtAl09}, mean estimated autosomal ancestry from a European ancestral group in African Americans has range 15-25\%, with standard deviation 8-15\%. Comparable values have been observed in subsequent studies~\citep{BrycEtAl15, BaharianEtAl16, MichelettiEtAl20}.

Because we treat the African-American population as a two-source group, we assume the African and European ancestry components sum to 1. As $\mathbb{V}[X]=\mathbb{V}[1-X]$ for a random variable $X$, we assume the two ancestry components have the same variance. Hence, to find parameter sets that give rise to admixture estimates that match those seen by~\cite{ChengEtAl09}, we search the parameter space for parameter sets that satisfy (i) the mean African ancestry, $\mathbb{E}[H_{1,14}]$, lies in $[0.75,0.85]$, and (ii) the standard deviation of the African ancestry, $\sqrt{\mathbb{V}[H_{1,14}]}$, lies in $[0.08,0.15]$.

We choose model parameters on a grid, and we then retain those sets of parameter values that satisfy the required conditions. For each parameter set that is retained, we calculate the mean, variance, and distribution of $Z_{1,14}$ and $Z_{2,14}$ by eqs.~\ref{eqMean}, \ref{eqVar}, and \ref{eqZDistribution}, respectively. We also compute the contributions of specific generations to the mean number of genealogical ancestors, following eq.~\ref{eqNonrecursive}. We characterize the properties of the parameter sets that we retain.

The analysis has one free parameter for epoch 1 (the European contribution, say, $s_{2,1}$); for epochs 2 and 3, it has three parameters each ($s_{1,7}$, $h_7$, $s_{2,7}$ and $s_{1,10}$, $h_{10}$, $s_{2,10}$), with two of three free to vary in each trio, as the trio necessarily sums to 1. We consider all possible points on a grid with increment $0.01$ for each parameter, enforcing an upper bound on the European contributions in all epochs due to the understanding that the African and African-American contributions predominate, an upper bound on the African contribution in epochs 2 and 3 due to comparatively low African immigration in these periods, and a lower bound on the African-American contribution in epochs 2 and 3 as a result of its equaling one minus the European and African contributions (Table \ref{table:S1constraints}).

\subsection*{Estimated model parameters}

\begin{figure}[tb]
\begin{center}
\centerline{
\includegraphics[width=1.3\textwidth]{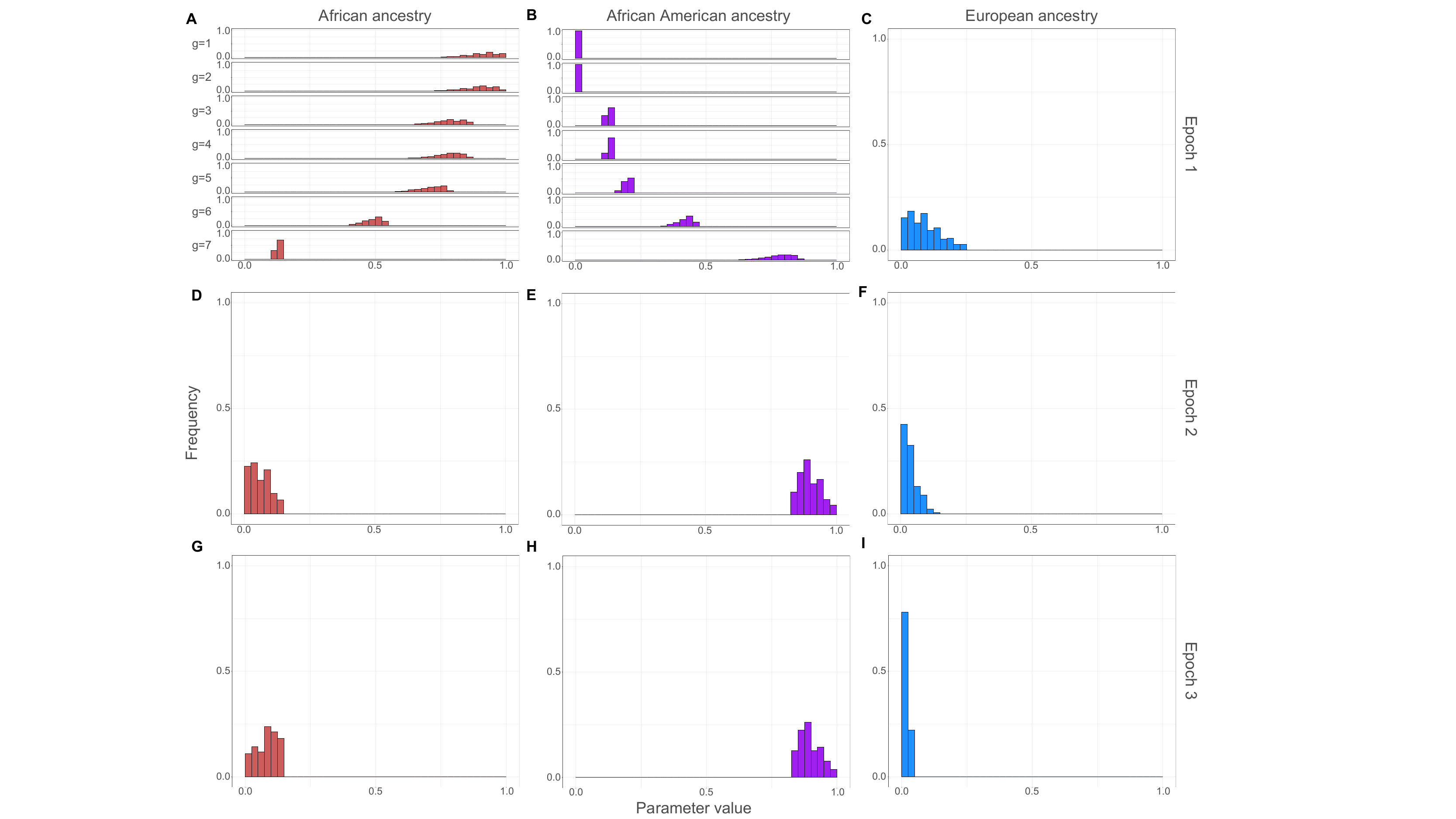}
}
\end{center}
\vspace{-0.7cm}
\caption{Distributions of generation-wise ancestry contributions estimated for African Americans. Generation-wise ancestry contributions are estimated for Africans, African Americans, and Europeans and European Americans. For each population in epochs 2 and 3, and for Europeans in epoch 1, the contribution from that population is assumed to be equal across generations within the epoch; for Africans and African Americans in epoch 1, the contribution changes across generations according to Table \ref{table:1cg}. The histograms are constructed from among accepted parameter sets that satisfied specified criteria. In epoch 1, the plots labeled with generation $g$ are the estimates of the parameters that contributed to births of individuals in generation $g$, representing $s_{1,g-1}$ and $h_{g-1}$. Parameter values are binned in intervals $[0,0.025], (0.025, 0.05], \ldots, (0.975, 1]$, half-open in all cases except the closed first bin.}\label{fig4}
\end{figure}

\noindent Distributions of the estimated model parameter sets that produce a mean and variance of African ancestry within permissible ranges appear in Figure \ref{fig4}, and they are summarized in Table \ref{table:2parameters}. In epoch 1, the generation-wise European ancestry contribution lies near the low end of the assumed range (Figure \ref{fig4}C), with a median of 0.08 (Table \ref{table:2parameters}). For this epoch, the African and African-American ancestry contributions are determined from demographic information and the European contribution (see Table \ref{table:1cg}); the estimated African contribution decreases from one generation to the next from the beginning to the end of the epoch (Figure \ref{fig4}A), and the African-American component increases (Figure \ref{fig4}B).

In epoch 2, the European contribution has median 0.03 (Table \ref{table:2parameters}), and the distribution of this contribution is concentrated at smaller values than in epoch 1 (Figure \ref{fig4}F). The African ancestry contribution is also small (Figure \ref{fig4}D), with median 0.06; most of the ancestry lies in the African-American component (Figure \ref{fig4}E).

Finally, in epoch 3, the European contribution decreases further to a median of 0.02 (Table \ref{table:2parameters}), with all the weight placed in the first two bins in Figure \ref{fig4}I. The African and African-American contribution components are similar to those seen in epoch 2 (Figure \ref{fig4}G,H), with a slight increase in the median African component (Table \ref{table:2parameters}).

\subsection*{Estimated numbers of genealogical ancestors}

\begin{figure}[tb]
\centerline{\includegraphics[width=0.5\textwidth]{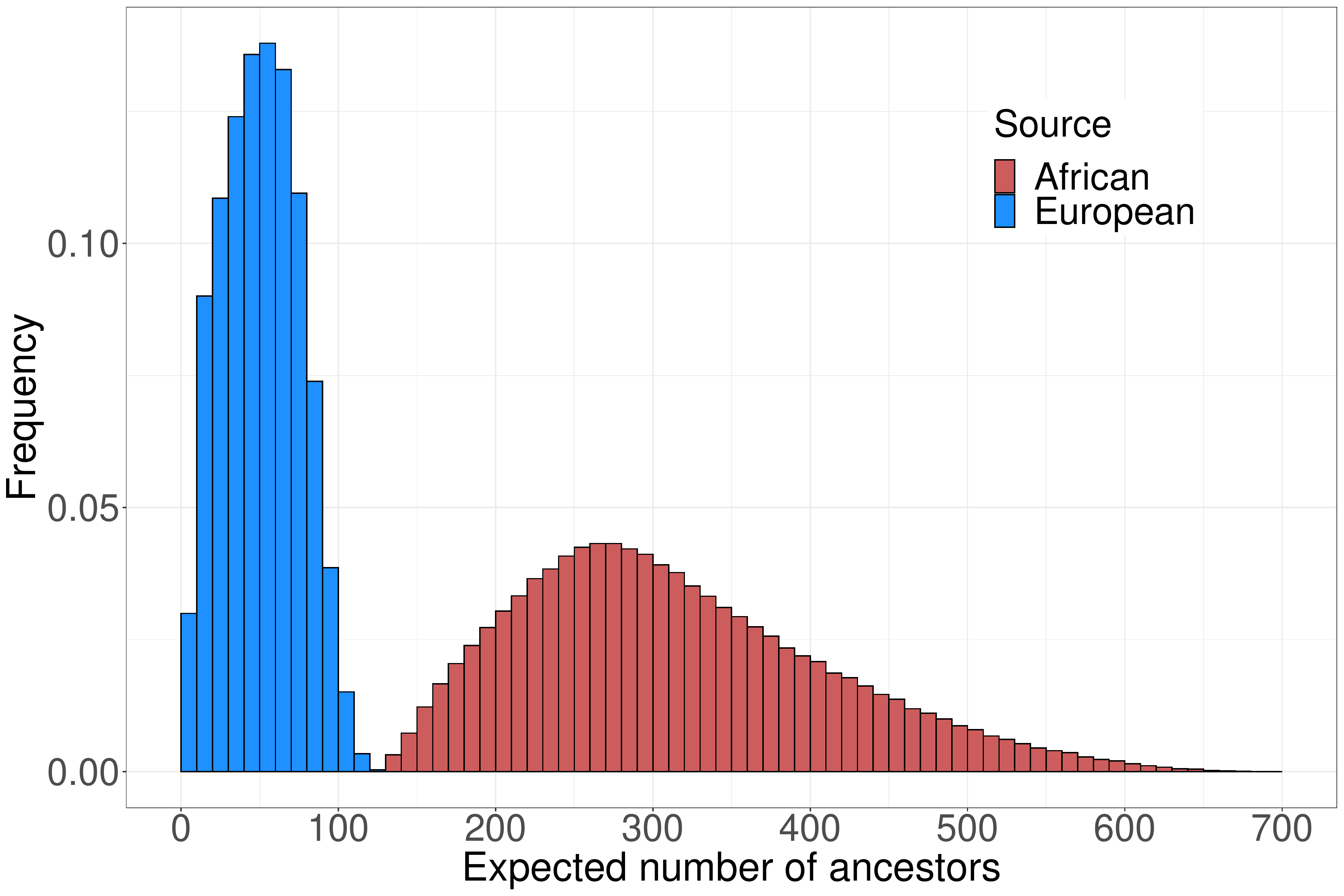}}
\vspace{0cm}
\caption{Distribution of the expectation of the numbers of African and European ancestors across accepted parameter sets. For each accepted set of parameter values, the expected number of African ancestors and the expected number of European ancestors are computed from eq.~\ref{eqMean}. Summaries of the figure appear in Table \ref{table:3results}.}\label{fig5}
\end{figure}

\noindent Each accepted parameter set generates values for the expected numbers of African and European genealogical ancestors, and the distributions of these quantities appear in Figure \ref{fig5} and Table \ref{table:3results}. The expected number of African ancestors has a mean of 314 and a median of 299, with an interquartile range from 240 to 376 and a minimum of 124 and maximum of 680 (Table \ref{table:3results}). The expected number of European ancestors is smaller and more concentrated, with mean 51, median 51, and interquartile range from 32 to 69; the minimum is 4 and the maximum is 125.

\begin{figure}[tb]
\centerline{\includegraphics[width=\textwidth]{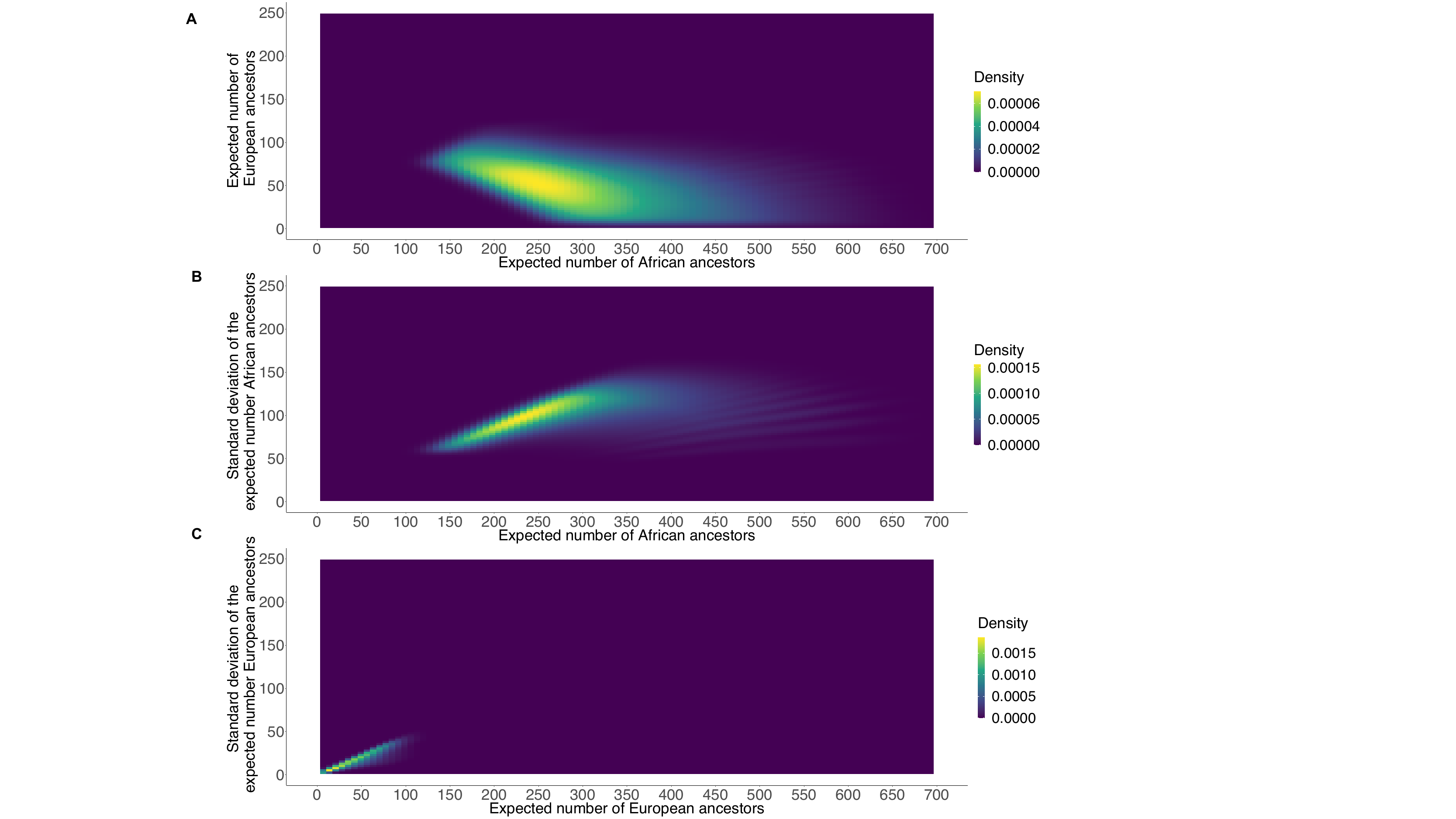}}
\vspace{0cm}
\caption{Joint distributions of the expectations and standard deviations of the numbers of African and European ancestors across accepted parameter sets. For each accepted set of parameter values, the expected number of African ancestors and the expected number of European ancestors are computed from eq.~\ref{eqMean}; the associated standard deviations are computed from eq.~\ref{eqVar}. (A) Expected number of European ancestors and expected number of African ancestors. (B) Standard deviation of the number of African ancestors and expected number of African ancestors. (C) Standard deviation of the number of European ancestors and expected number of European ancestors.}\label{fig6}
\end{figure}

Considering the expected numbers of African and European ancestors jointly, we observe that across accepted parameter sets, they are negatively correlated ($r=-0.455$, Figure \ref{fig6}A). For both Africans and Europeans, the standard deviation of the number of ancestors increases with the associated expectation ($r=0.434$ for Africans, Figure \ref{fig6}B; $r=0.900$ for Europeans, Figure \ref{fig6}C).

\begin{figure}[tb]
\centerline{\includegraphics[width=0.75\textwidth]{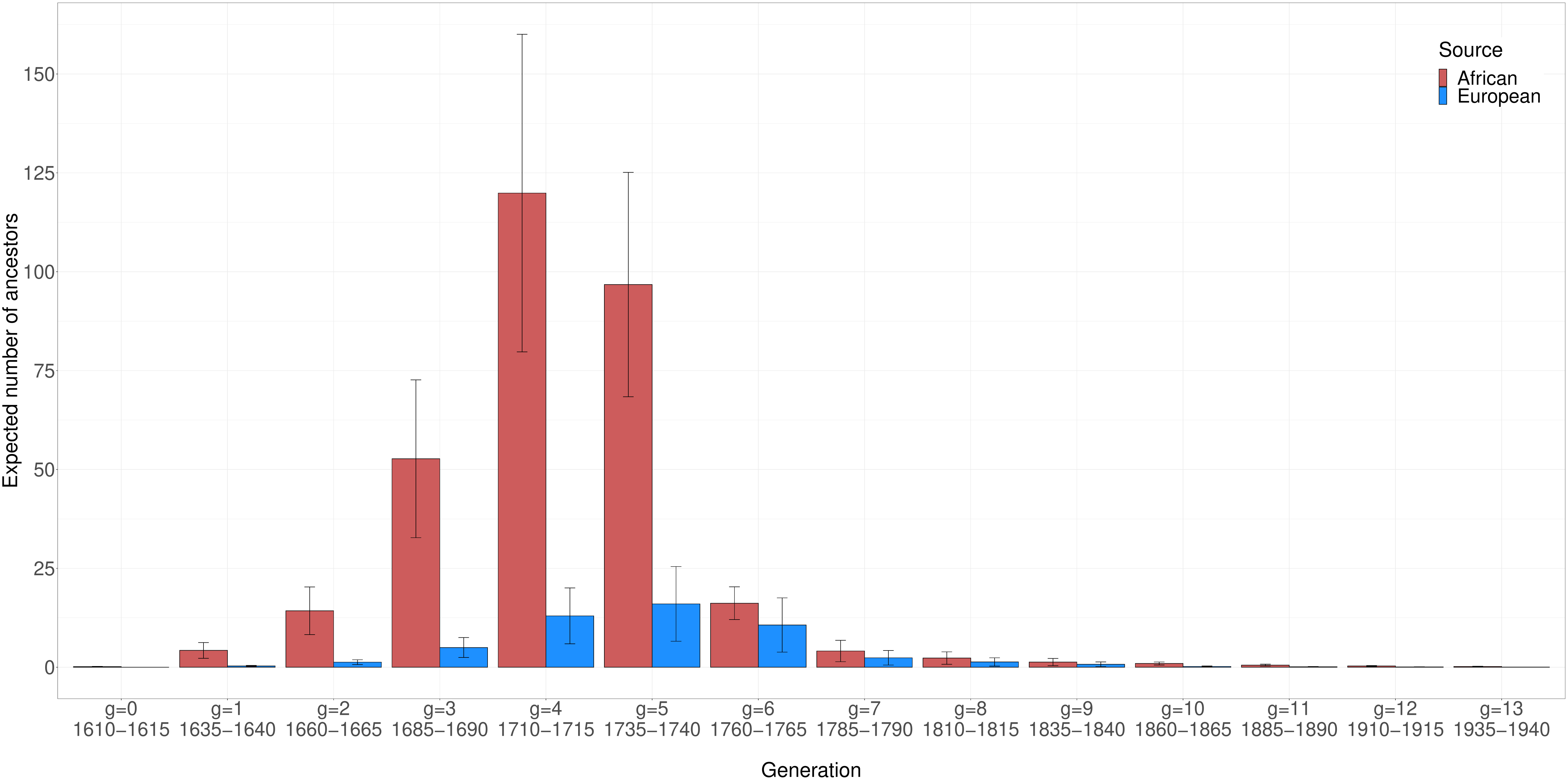}}
\vspace{0cm}
\caption{Generation-specific expectations of the numbers of African and European ancestors across accepted parameter sets. For each accepted set of parameter values, the generation-specific expected number of African ancestors and the generation-specific expected number of European ancestors are computed from eq.~\ref{eqNonrecursive}. The height of a bar represents the mean across accepted parameter sets of the generation-specific expected number of ancestors, and the error bars represent standard deviations.}
\label{fig7}
\end{figure}

Separating the African and European ancestors by their generational timing  (Figure \ref{fig7} and Table \ref{table:S2generations}), we see that the greatest numbers trace to epoch 1, particularly generations 3-5 for Africans (1685-1740) and 4-6 for Europeans (1710-1765). Nonzero values for both quantities continue, decreasing to small values in the most recent generations.

\subsection*{Probability of at least one genealogical ancestor}

\noindent Applying the estimated means for the admixture parameters, we used eq.~\ref{eq:U} to evaluate the probability for each generation that an African-American individual has at least one African genealogical ancestor in that generation, and the corresponding probability that an African-American individual has at least one European genealogical ancestor.

\begin{figure}[tb]
\centerline{\includegraphics[width=0.75\textwidth]{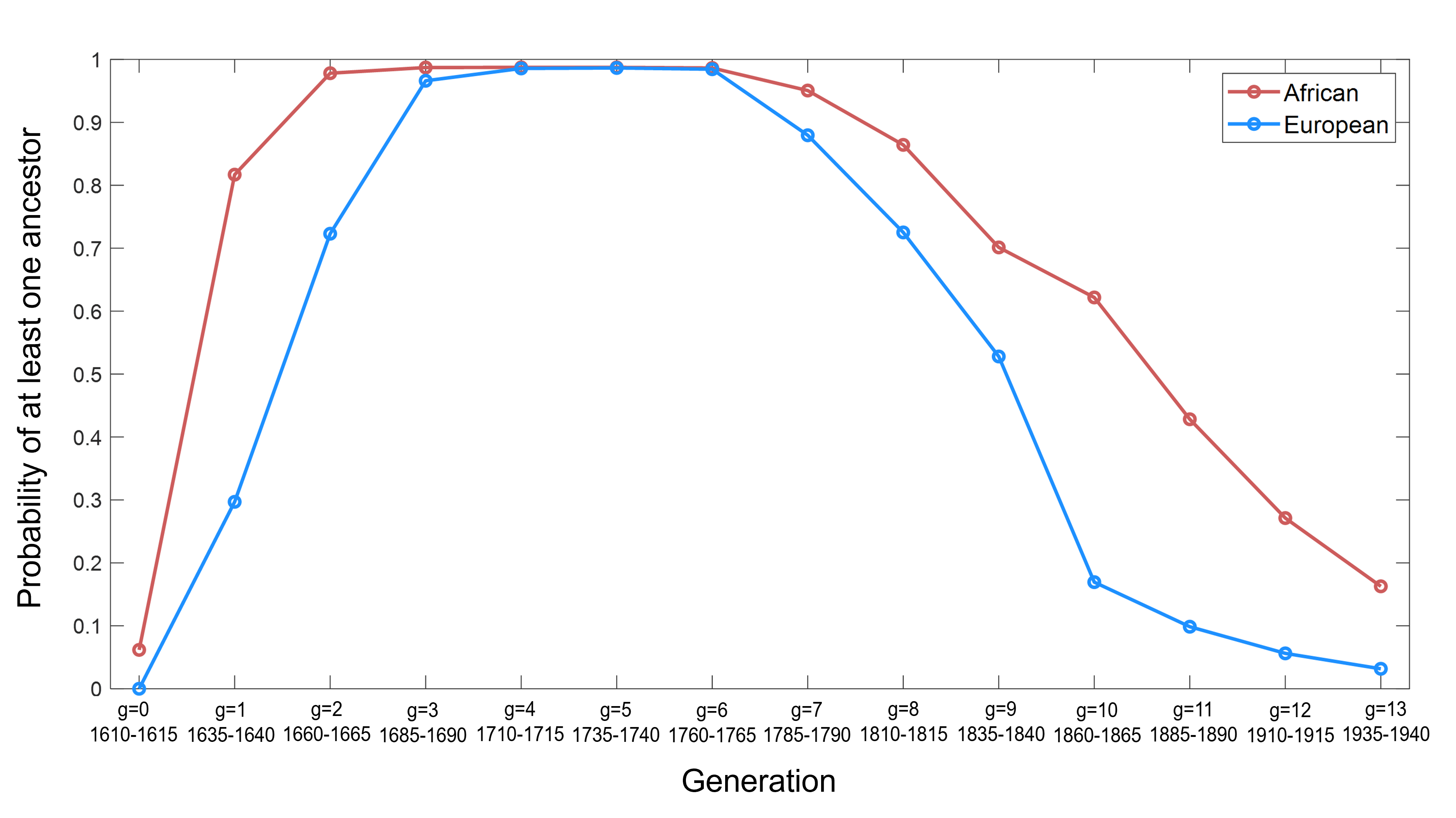}}
\vspace{0cm}
\caption{The probability of having at least one ancestor from a source population in a specified generation.  Considering the means among accepted parameter sets, $(s_{1,0},s_{1,1}, \ldots ,s_{1,13}) = (1,0.896,0.783,0.779,0.713,0.490,0.129,0.061,0.061,0.061,0.085,0.085,0.085,0.085)$, $(h_0,h_1, \ldots ,h_{13}) = (0,0.015,0.127,0.132,0.198,0.421,0.781,0.902,0.902,0.902,0.899,0.899,0.899,0.899)$,
and $(s_{2,0},s_{2,1}, \ldots ,s_{2,13})=(0,0.089,0.089,0.089,0.089,0.089,0.089, 0.037,0.037,0.037, 0.016,0.016,0.016,0.016)$ (Tables \ref{table:1cg} and \ref{table:2parameters}), the generation-specific probabilities of at least one African ancestor and at least one European ancestor are computed from eq.~\ref{eq:U}.}
\label{fig8}
\end{figure}

Figure \ref{fig8} plots this probability. For African ancestors, the probability is small for generation 0, increasing for generations 2-6, and then decreasing. For each of generations 2-6, the probability exceeds $0.975$ that a random African American has at least one African ancestor in that generation (Table \ref{table:S3probability}). In other words, the probability is near 1 that in each of generations 3-7, the offspring generations of generations 2-6, at least one individual in a random genealogy has an African parent.

In Figure \ref{fig8}, in each generation, the probability of at least one European ancestor has a similar pattern, with its largest values in generations 4-6. It remains above 0.5 in each of generations 7-9, and it is substantially lower in generations 10-13. In each generation, the probability of at least one European ancestor is smaller than the corresponding probability of at least one African ancestor.


\section*{Discussion}

\noindent Under models of admixture, we have evaluated the numbers of genealogical lines that trace to particular source populations. The results provide a new perspective on admixture models, focusing on properties of individual genealogies. We have applied this perspective to the case of African Americans, finding that under a model calibrated by demographic data on African-American admixture, a random African-American genealogy traced back in time from birth in 1960-1965 reaches a mean of 314 African individuals and 51 European or European-American individuals.

\subsection*{Admixture models}

\noindent Our approach builds on mechanistic admixture models that have characterized the distribution of admixture levels over time as a function of model parameters. The quantities that we examine---properties of the distributions of the number of ancestors from the source populations---are obtained as functions of model parameters in a manner similar to the computation of the distributions of admixture levels. Estimated individual-level genomic admixture fractions are used to calibrate the models, from which aspects of the numbers of ancestors are calculated in terms of model parameters.

In standard coalescent approaches, the genealogy of a single locus is traced among many individuals back to a common ancestor---disregarding diploid pedigrees. Recent genealogical analyses have sought to also include pedigrees and to examine stochastic processes involving gene lineages on those pedigrees~\citep{WollenbergAndAvise98, WakeleyEtAl12, Campbell15, WakeleyEtAl16, WiltonEtAl17, SeversonEtAl19}. Such studies often analyze properties of genealogical rather than genetic ancestry, using theoretical and simulation-based approaches~\citep{RohdeEtAl04, MatsenAndEvans08, Lachance09, GravelAndSteel15, KelleherEtAl16, EdgeAndCoop20}. Our investigation of genealogical lines in admixed populations continues a series that provides a basis for investigating admixed biparental genealogies in the most recent generations~\citep{VerduAndRosenberg11, Gravel12, GoldbergEtAl14, GoldbergAndRosenberg15, GoldbergEtAl20, KimEtAl21}.

\subsection*{African-American demographic history}

\noindent The results provide insight into African-American history. First, the model suggests that patterns seen in African-American genetic ancestry correspond to a mean of $0.089$ for the generation-wise European ancestry component in epoch 1, $0.037$ in epoch 2, and $0.016$ in epoch 3 (Table \ref{table:2parameters}). These values have comparable magnitude to values in other studies that have estimated similar quantities, but without a 3-epoch perspective~\citep{GlassAndLi53, Gross18}. The European ancestry parameter decreases from the initial period through the last generations of enslavement, decreasing again after the end of slavery.

We estimate that a random African American born during 1960-1965 has a mean of 314 African ancestors and 51 European and European-American ancestors (Figure \ref{fig5} and Table \ref{table:3results}). The model finds that most genealogical lines trace back through African-American ancestors for several generations; at that point, the number of African-American ancestors is large, and some have African parents, European parents, or both. Most ancestors from the source populations, both African and European, appear in generations 3-6, 1685-1765 (Figure \ref{fig7}), with near 100 African ancestors each in generations 4 and 5 (Table \ref{table:S2generations}). As a genealogy proceeds back in time, for those genealogical lines that are not from the source populations, the number of lines doubles each generation, potentially driving the temporal maximum for the number of genealogical ancestors early in the history of the African-American population. In the early generations, the number of African parents is high relative to African-American parents, so that large numbers of African ancestors accumulate in a pedigree in those generations; in generations after generation 6, the number of African-American parents relative to African parents is high enough that fewer Africans appear. Interestingly, the peak importation of enslaved individuals did not occur until later in the 1700s than the African-ancestor peak~\citep[][p.~200]{EltisAndRichardson10}; by the time of the importation peak, the fraction of parents of a generation's offspring who were African-American rather than African was already relatively high~\citep{Hacker20}.

The ancestor counts can be approached by a focus on the earliest African ancestor: for a random African American, what is the distribution of the generation in which the earliest African ancestor lived? In Figure \ref{fig8} and Table \ref{table:S3probability}, for each of generations 2-6, the probability exceeds 97\% that a random African American contains at least one African ancestor in that generation. In other words, the probability exceeds 97\% that in each of generations 3-7, the offspring generations of generations 2-6, at least one individual in a random genealogy is an African American with an African parent. Considering the earliest of these generations, under the model, a typical African American born in 1960-1965 likely has at least one ancestor from generation 4 (1710-1715) who was an African American with an African parent, and it is also likely that such an individual has at least one African-American ancestor from generation 3 as well (1685-1690).

For European ancestors, we find that under the model, the probability is high ($>$96\%) that a random African-American individual has at least one European ancestor in each of generations 3-6, the parents of generations 4-7 (Figure \ref{fig8}). Although fewer European ancestors are present in generations 7-9 than 3-6, the probability of a European ancestor exceeds 50\% in each of generations 7-9. In other words, for example, the probability is above 50\% that a random African-American individual has a European ancestor born in generation 9 (1835-1840).

Among the parameter estimates, $0.085$ for African ancestry in epoch 3 is potentially misaligned with historical information; this value is large given low levels of African immigration during the period~\citep{Reimers05, Gates09, Berlin10}. This estimate may reflect any of a number of phenomena. First, individuals from the Caribbean potentially have high African ancestry fractions~\citep{MichelettiEtAl20, MathiasEtAl16, AdhikariEtAl17}; some of the apparent African immigration detected in epoch 3 might, instead, be misattributed immigration from the Caribbean, a source of more migrants than Africa during the period, though still a small number relative to the resident African-American population~\citep{Henke01, Reimers05, Berlin10}. Second, the African-American and African ancestry components are difficult to disentangle; that the admixed African-American population has greater genetic similarity to the African than to the European population decreases identifiability for the African and African-American components. Indeed these components are negatively correlated across accepted parameter sets (Table \ref{table:S4correlations}), and their levels of uncertainty in epoch 3 exceed that of the European component (Figure \ref{fig4}G-I). An overestimation of the African ancestry component in epoch 3---when the true African ancestry traces to earlier epochs---means that the model may be placing larger fractions of individual pedigrees in the African source population in recent generations than is warranted. To produce the desired mean African ancestry level, one African ancestor in epoch 3 contributes the same amount of African ancestry as multiple African ancestors from earlier epochs. Hence, if the African ancestry component in epoch 3 is an overestimate of the true value, then the model may be undercounting the true number of African ancestors --- so that a count of 314 African ancestors may in fact \emph{underestimate} the true count.

\subsection*{Interpretation in relation to a single African-American genealogy}

\noindent As limitations of African-American genealogical research impede the use of documentary evidence to count genealogical lineages that reach individual African and European ancestors in genealogies of specific individuals~\citep{Gates09, Swarns12, Nelson16}, our claim that a random African American born during 1960-1965 has a mean of 314 African and 51 European ancestors provides information that extends beyond what can typically be documented in individual genealogies. To illustrate the meaning of the results, we examine them in the context of a single specific genealogy.

Consider a genealogical study \citep{Swarns12} of a prominent African American: Michelle Obama, born in 1964, corresponding to generation 14 of our model. Her family history has many features typical of African-American genealogies, so that we can treat it as an instance of a random genealogy. The genealogy has 2 African-American parents, 4 African-American grandparents, and 8 African-American great-grandparents, and 10 known African-American great-great-grandparents; no evidence suggests that the other 6 great-great-grandparents are not African Americans. In the great-great-great-grandparental generation (generation 9 in our model), one European is identified, Charles Shields (born 1839), the father of African-American great-great-grandparent Dolphus Shields born circa 1859, with enslaved African-American mother Melvinia Shields (born c.~1844),

In one of the most extensively investigated African-American genealogies, in tracing back 5 generations (to generation 9 in our model), 1 specific named European is reached. From family photographs and oral histories, at least 3 other lineages likely terminate in a European in that generation or the one that precedes it (the James Preston Johnson, Phoebe Moten, and Jim Jumper lineages). No African ancestors are identifiable by name.

Michelle Obama's ancestors of the last 2-3 generations (generations 11-12) were part of a migration of millions of African Americans from the American South to northern cities~\citep{Lemann91, Berlin10, Wilkerson10}. Her ancestors 3-4 generations ago (generations 10-11) were African Americans living throughout the American South. The large number of southern locations from which they arrived in her home city of Chicago suggests that they can be viewed as a random sample from the region. Her ancestors in the fourth generation back from the present (generation 10) primarily included enslaved individuals and some free African Americans prior to 1865. The fifth generation back (generation 9) includes the likely most recent European appearance in a genealogy that consisted in that generation primarily of enslaved African Americans. Note that generation 9 is precisely the most recent generation identified in Figure \ref{fig8} during which the probability of a European ancestor exceeds 50\%.

The small number of African and European ancestors that can be named in a genealogically typical African American --- 1 European and 0 Africans --- can be compared with the estimate of the much larger actual number of ancestors. As the numbers of African and European ancestors in the two most recent generations (generations 12 and 13) are small in the model (Figure \ref{fig7}), our estimate of 314 African and 51 European ancestors approximately corresponds to a claim that for an African American such as Michelle Obama with 4 African-American grandparents, each of those grandparents has a mean of perhaps $314/4 = 78.5$ African and $51/4 = 12.75$ European ancestors.

In an additional interpretation of the African ancestors, forced voyages of enslaved migrants from Africa to the North American mainland had a fatality rate of $\sim$12-29\%~\citep[][p.~167]{EltisAndRichardson10}. Under the model, if it is assumed that almost all the African ancestors before 1808 were enslaved migrants and that no ancestor is an ancestor by multiple paths through a pedigree, then a random African American born in 1960-1965 is descended from, on average, $\sim$300 separate survivors of these journeys. For the European ancestors, although genetic studies have found that African Americans have $\sim$20\% European ancestry on average, the equivalent of more than one European great-grandparent (12.5\% ancestry), African Americans whose recent ancestors are all African Americans might have no European ancestors who are specifically known to them: for Michelle Obama, the most recent European ancestor was discovered by a genealogist~\citep{Swarns12}. Our estimate of a mean of 51 European ancestors amounts to a claim that for a typical African-American genealogy of a person born from 1960-1965, the generations since the founding of the population contain a mean of 51 separate mating events between a European or European American and an African or African American.

\subsection*{Limitations}

\noindent Our analyses of African-American demographic history make use of empirical estimates of admixture levels together with information on the demographics of enslavement \citep{Hacker20}. However, we note that it does not consider a variety of known phenomena of African-American demographic history.

First, we have treated the African-American population as the outcome of admixture only between African and European sources, and we have not considered Native-American or other sources. Genomic studies generally find that the Native-American contribution is small \citep{BrycEtAl15, BaharianEtAl16}, 3\% or less, and that that the distribution across African Americans of the Native-American ancestry component is more difficult to accurately estimate than the African and European contributions. With a model that includes the Native-American contributions as a third source, the distribution of the number of Native-American ancestors could potentially be estimated.

We also have not considered variation in African and European admixture across the United States. To calibrate the model, we chose a range of admixture estimates for African and European admixture, based on studies in many locations. Parameter estimates for our model of African-American admixture history represent a composite of many subpopulations; in some regions, the numbers of African and European genealogical ancestors might differ from these composite values.

Finally, we have assumed that ancestral individuals do not appear in a genealogy on multiple paths. Among millions of African Americans over American history, multiple genealogical lineages might reach the same ancestor; we have assumed that such ancestor-sharing events are rare in individual genealogies. The number of enslaved African migrants brought to the United States has been estimated near $\sim$400,000 prior to 1825~\citep[][p.~200]{EltisAndRichardson10}. With 314 African ancestors for a random individual, it is possible that two or more genealogical lines reach the same individual among the $\sim$400,000. Duplication of lines is most likely in the early history of the admixed population, in which the population had the smallest size, and in which many of the ancestors are assigned (Figure \ref{fig7}). However, as 314 is small in relation to 400,000, any possible overestimation of the number of ancestors due to these duplications is likely to be relatively small.

\subsection*{Conclusions}

\noindent This study introduces new quantities into the genetic study of admixed populations, namely the numbers of genealogical ancestors in an individual genealogy who were members of the source populations. We have shown how to calculate these quantities from a mechanistic model of ancestry whose parameters can be estimated from admixture levels in an admixed population. The approach yields new information for understanding the history of admixed populations, and in the case of African Americans, it sheds light on an admixture process many of whose genealogical and demographic aspects are difficult to access by other means.

\fontsize{9}{18}\selectfont
\vskip .7cm
\noindent
{\bf Acknowledgments.} We thank Janina Jeff, Tina Lasisi, John Thornton, and Naomi Zack for helpful discussions. Support was provided by National Science Foundation grant BCS-2116322, a National Science Foundation Postdoctoral Research Fellowship in Biology, a Council for Higher Education of Israel Scholarship for Outstanding Postdoctoral Fellows in Data Science, and the Stanford Center for Computational, Evolutionary, and Human Genomics.

\vskip .7cm
\bibliography{genealogical3}

\fontsize{10.5}{21}\selectfont

\clearpage
\begin{table}[tb]
    \centering
    \begin{tabular}{lccl}
    \hline
        Generation $g$ & Birth year & Epoch & $c_{g-1}$ \\
\hline
        1  & 1635-1640 & 1 & 1 \\
        2  & 1660-1665 & 1 & 0.9835 \\
        3  & 1685-1690 & 1 & 0.8602 \\
        4  & 1710-1715 & 1 & 0.8551 \\
        5  & 1735-1740 & 1 & 0.7826 \\
        6  & 1760-1765 & 1 & 0.5380 \\
        7  & 1785-1790 & 1 & 0.1418 \\
        8  & 1810-1815 & 2 & - \\ 
        9  & 1835-1840 & 2 & - \\ 
        10 & 1860-1865 & 2 & - \\ 
        11 & 1885-1890 & 3 & - \\
        12 & 1910-1915 & 3 & - \\
        13 & 1935-1940 & 3 & - \\
        14 & 1960-1965 & 3 & - \\
        \hline
    \end{tabular}
    \caption{Parametrizing a historically informed model. For all generations $g$ in epoch 1 ($g$ from 1 to 7), the quantity $c_{g-1}=s_{1,g-1}/(1-s_{2,g-1})$ denotes, for individuals born in the African-American population in generation $g$, the fraction of their non-European parents who are African arrivals to North America rather than African-American residents. In our model, we inserted numerical values for this quantity estimated based on demographic data.}
    \label{table:1cg}
\end{table}

\clearpage
\begin{table}[tb]
    \centering
    \begin{tabular}{llccccccc}
    \hline
Epoch   & Population                   & Mean  & Standard  & Minimum & 1st      & Median & 3rd      & Maximum  \\
        &                              &       & deviation &         & quartile &        & quartile &          \\
\hline
Epoch 1 & European ($s_{2,1})$         & 0.089 & 0.061     & 0       & 0.04     & 0.08   & 0.13     & 0.25 \\
Epoch 2 & African ($s_{1,7})$          & 0.061 & 0.040     & 0       & 0.03     & 0.06   & 0.09     & 0.15 \\
        & African-American ($h_{7})$   & 0.902 & 0.039     & 0.85    & 0.87     & 0.90   & 0.93     & 1.00 \\
        & European ($s_{2,7})$         & 0.037 & 0.030     & 0       & 0.01     & 0.03   & 0.05     & 0.15 \\
Epoch 3 & African ($s_{1,10})$         & 0.085 & 0.041     & 0       & 0.05     & 0.09   & 0.12     & 0.15 \\
        & African-American ($h_{10})$  & 0.899 & 0.039     & 0.85    & 0.87     & 0.89   & 0.93     & 0.99 \\
        & European ($s_{2,10})$        & 0.016 & 0.010     & 0       & 0.01     & 0.02   & 0.02     & 0.03 \\
\hline
    \end{tabular}
    \caption{Estimated model parameters for a 3-epoch model of African-American demographic history. The table summarizes the parameter sets that produce permissible values for the expectation and variance of $H_{1,14}$, the African ancestry fraction in generation 14. Note that in epoch 1, the African and African-American parameter values are generation-specific, set according to the values in Table \ref{table:1cg} rather than estimated. The table is based on $45,189$ accepted parameter sets, $\sim 9\%$ of the $480,896$ sets examined.}
    \label{table:2parameters}
\end{table}

\clearpage
\begin{table}[tb]
    \centering
    \begin{tabular}{lccccccc}
    \hline
Quantity & Mean & Standard  & Minimum & 1st      & Median & 3rd      & Maximum \\
         &      & deviation &         & quartile &        & quartile &         \\
\hline
African ancestors    & 314   & 103  & 124 & 240 & 299 & 376 & 680 \\
European ancestors   & 51    & 19   & 4   & 32  & 51  & 69  & 125 \\
\hline
    \end{tabular}
    \caption{Summary statistics for the expected numbers of African and European ancestors for a random individual from the African-American population ($\mathbb{E}[Z_{1,14}]$ and $\mathbb{E}[Z_{2,14}]$). The estimates consider random individuals in the 1960-1965 birth cohort, assumed to be generation $g=14$ in a 3-epoch model. The quantities in the table summarize results plotted in Figure \ref{fig5}.}
    \label{table:3results}
\end{table}

\clearpage

\clearpage

\clearpage

\clearpage

\clearpage

\clearpage

\clearpage

\clearpage

\setcounter{figure}{0}
\setcounter{table}{0}
\renewcommand{\thetable}{S\arabic{table}}
\renewcommand{\thefigure}{S\arabic{figure}}

\clearpage
\begin{table}[tb]
    \centering
    \begin{tabular}{|ll|cc|}
    \hline
Epoch   & Population                   & Minimum  & Maximum \\
\hline
Epoch 1 & European ($s_{2,1})$         & 0     & 0.25  \\
Epoch 2 & African ($s_{1,7})$          & 0     & 0.15  \\
        & African-American ($h_{7})$   & 0.85  & 1.00  \\
        & European ($s_{2,7})$         & 0     & 0.15  \\
Epoch 3 & African ($s_{1,10})$         & 0     & 0.15  \\
        & African-American ($h_{10})$  & 0.85  & 1.00  \\
        & European ($s_{2,10})$        & 0     & 0.15  \\
\hline
    \end{tabular}
    \caption{Ranges assumed for model parameters for a 3-epoch model of African-American demographic history. Note that in epoch 1, the African and African-American parameter values are generation-specific, set at $c_{g-1}(1-s_{2,g-1})$ and $(1-c_{g-1})(1-s_{2,g-1})$, respectively, according to the values in Table \ref{table:1cg}. With an increment of 0.01 in the parameters, the total number of parameter sets considered is 480,896, the product of 26 choices for epoch 1 and $(16)(17)/2=136$ each for epochs 2 and 3.}
    \label{table:S1constraints}
\end{table}

\clearpage
\begin{table}[tb]
    \centering
    \begin{tabular}{|c|c|cc|cc|}
    \hline
               &            & \multicolumn{2}{c|}{African ancestors} & \multicolumn{2}{c|}{European ancestors} \\ \cline{3-6}
               &            & Mean & Standard  & Mean & Standard  \\
Generation $g$ & Birth year &      & deviation &      & deviation \\ \hline
        0  & 1610-1615 &   0.143 &   0.067 &      - &      - \\
        1  & 1635-1640 &   4.249 &   1.987 &  0.317 &  0.144 \\
        2  & 1660-1665 &  14.267 &   6.036 &  1.277 &  0.609 \\
        3  & 1685-1690 &  52.705 &  19.953 &  4.977 &  2.524 \\
        4  & 1710-1715 & 119.896 &  40.146 & 12.983 &  7.075 \\
        5  & 1735-1740 &  96.762 &  28.370 & 16.007 &  9.444 \\
        6  & 1760-1765 &  16.181 &   4.145 & 10.672 &  6.853 \\
        7  & 1785-1790 &   4.098 &   2.708 &  2.379 &  1.837 \\
        8  & 1810-1815 &   2.314 &   1.572 &  1.333 &  1.044 \\
        9  & 1835-1840 &   1.308 &   0.914 &  0.748 &  0.595 \\
        10 & 1860-1865 &   0.936 &   0.386 &  0.189 &  0.116 \\
        11 & 1885-1890 &   0.530 &   0.230 &  0.104 &  0.063 \\
        12 & 1910-1915 &   0.301 &   0.137 &  0.058 &  0.035 \\
        13 & 1935-1940 &   0.171 &   0.082 &  0.032 &  0.019 \\ \hline
     Total &   -       & 313.859 & 102.769 & 51.076 & 18.736 \\ \hline
    \end{tabular}
    \caption{Generation-specific expectations of the numbers of African and European ancestors across accepted parameter sets. The table shows the values plotted in Figure \ref{fig7}.}
    \label{table:S2generations}
\end{table}

\clearpage
\begin{table}[tb]
\centering
\begin{tabular}{|cc|cc|}
 \hline
               &            & \multicolumn{2}{c|}{Probability of at least one ancestor} \\
Generation $g$ & Birth year & African & European \\
 \hline
0  & 1610-1615 & 0.0618 & 0.0000 \\
1  & 1635-1640 & 0.8169 & 0.2970 \\
2  & 1660-1665 & 0.9781 & 0.7230 \\
3  & 1685-1690 & 0.9872 & 0.9661 \\
4  & 1710-1715 & 0.9874 & 0.9857 \\
5  & 1735-1740 & 0.9874 & 0.9865 \\
6  & 1760-1765 & 0.9864 & 0.9844 \\
7  & 1785-1790 & 0.9506 & 0.8795 \\
8  & 1810-1815 & 0.8642 & 0.7251 \\
9  & 1835-1840 & 0.7013 & 0.5280 \\
10 & 1860-1865 & 0.6218 & 0.1694 \\
11 & 1885-1890 & 0.4283 & 0.0986 \\
12 & 1910-1915 & 0.2713 & 0.0563 \\
13 & 1935-1940 & 0.1628 & 0.0317 \\
 \hline
\end{tabular}
\caption{The probability of having at least one ancestor from a source population in a specified generation. The table shows the values plotted in Figure \ref{fig8}.}
\label{table:S3probability}
\end{table}

\clearpage
\begin{table}[tb]
    \centering
    \begin{tabular}{|c|c|c|rrr|rrr|}
    \hline
      &           & Epoch 1 & \multicolumn{3}{c|}{Epoch 2} & \multicolumn{3}{c|}{Epoch 3} \\ \cline{3-9}
Epoch & Parameter & $s_{2,1}$ & $s_{1,7}$ & $h_7$ & $s_{2,7}$ & $s_{1,10}$ & $h_{10}$ & $s_{2,10}$ \\ \hline
3 & $s_{2,10}$    & &       &        &        &        &        &        \\
3 & $h_{10}$      & &       &        &        &        &        &  0.095 \\
3 & $s_{1,10}$    & &       &        &        &        & -0.972 & -0.328 \\ \hline
2 & $s_{2,7}$     & &       &        &        &  0.281 & -0.217 & -0.318 \\
2 & $h_{7}$       & &       &        & -0.361 & -0.061 &  0.046 &  0.073 \\
2 & $s_{1,7}$     & &       & -0.708 & -0.404 & -0.153 &  0.118 &  0.169 \\ \hline
1 & $s_{2,1}$     & & 0.336 & -0.013 & -0.427 &  0.430 & -0.352 & -0.402 \\
\hline
    \end{tabular}
    \caption{Pearson correlations of estimated parameters. The correlations are computed across all accepted parameter sets.}
    \label{table:S4correlations}
\end{table}

\end{document}